\documentclass[aps,prb,preprint,groupedaddress]{revtex4}

\usepackage{graphicx}
\usepackage{dcolumn}
\usepackage{bm}

\begin{document}

\title{Effect of DC electric field on longitudinal resistance of two
dimensional electrons in a magnetic field. }

\author{Jing-qiao Zhang}
\author{Sergey Vitkalov}
\email[Corresponding author: ]{vitkalov@sci.ccny.cuny.edu}
\affiliation{Physics Department, City College of the City University of New York, New York 10031, USA}
\author{ A. A. Bykov, A. K. Kalagin and A. K. Bakarov }
\affiliation{Institute of Semiconductor Physics, 630090 Novosibirsk, Russia}

\date{\today}

\begin{abstract}

The effect of a DC electric field on the longitudinal resistance of
highly mobile two dimensional electrons in heavily doped GaAs quantum
wells is studied at different magnetic fields and temperatures. Strong
suppression of the resistance by the electric field is observed in
magnetic fields at which the Landau quantization of electron motion
occurs. The phenomenon survives at high temperature where Shubnikov de
Haas oscillations are absent. The scale of the electric fields essential
for the effect is found to be proportional to temperature in the low
temperature limit.  We suggest that the strong reduction of the
longitudinal resistance is the result of a nontrivial change in the
distribution function of 2D electrons induced by the DC electric field. 
Comparison of the data with recent theory yields the inelastic electron-electon 
scattering time $\tau_{in}$ and the quantum scattering
time $\tau_q$ of 2D electrons at high temperatures, a regime where previous
methods were not successful.   
    
\end{abstract}

\pacs{}

\maketitle


The nonlinear properties of highly mobile two dimensional electrons
in AlGaAs/GaAs heterojunctions is a subject of considerable current
interest. Strong oscillations of the longitudinal resistance induced by
microwave radiation have been found at magnetic fields which satisfy the
condition 
$\omega=n \times \omega_c$, where $\omega$ is the microwave frequency, 
$\omega_c$ is cyclotron frequency and $n$=1,2...\cite{zudov,engel}.   At high levels of the
microwave excitations the minima of the oscillations can reach values
close to zero \cite{mani,zudov2,dorozh1,willett}.  This so-called
zero resistance state (ZRS) has stimulated extensive theoretical interest
\cite{andreev,durst,anderson,shi,vavilov,dmitriev}.  At higher magnetic
field $\omega_c > \omega$ a considerable decrease of magnetoresistance
with microwave power is found \cite{engel,dorozh1,willett} which has
been  attributed to intra-Landau-level transitions\cite{dorozh2}.

Another interesting nonlinear phenomenon has been observed in response to
DC electric field \cite{yang,bykov1}.  Oscillations of the longitudinal
resistance, which are periodic in inverse magnetic field, have been found
at DC biases, satisfying the condition $n \times \hbar \omega_c=2 R_c
E_H$, where $R_c$ is the Larmor radius of electrons at the Fermi level and
$E_H$ is the Hall electric field induced by the DC bias in the magnetic
field. The effect has been attributed to "horizontal" Landau-Zener
tunneling between Landau levels, tilted by the Hall electric field $E_H$ 
\cite{yang}.

In this paper we report a new phenomenon.  We have observed a strong
reduction of the 2D longitudinal resistance induced by DC electric
field $E_{dc}$ which is substantially smaller that required for the
"horizontal" electron transitions between Landau levels\cite{yang,bykov1}.
In contrast to the inter Landau level scattering, the observed effect
depends strongly on temperature. We suggest that the phenomenon is due to
a substantial and nontrivial deviation of the electron distribution
function from equilibrium induced by the DC electric field $E_{dc}$.  We
find reasonable agreement between our results and a recent theory that
considers such an effect in the high temperature limit\cite{dmitriev}.   

Our samples were cleaved from a wafer of high-mobility GaAs quantum well
grown by molecular beam epitaxy on semi-insulating (001) GaAs
substrates. The width of the GaAs quantum well was 13 nm. AlAs/GaAs
type-II superlattices served as barriers, making possible 
a high-mobility 2D electron gas with high electron density \cite{fried}.
Two samples (N1 and N2) were studied with electron density $n_1$ = 1.22 $\times
10^{16}$ m$^{-2}$, $n_2$ = 0.84 $\times 10^{16}$ m$^{-2}$,and
mobility $\mu_1$= 93 m$^2$/Vs, $\mu_2$= 68 m$^2$/Vs at T=2.7K.
Measurements were carried out between T=1.8 K and T=77 K in magnetic
field up to 3.2 T  on $d$=50 $\mu m$ wide Hall bars with a distance of 250
$\mu m$ between potential contacts. The longitudinal resistance was
measured using a current of 0.5 $\mu$A at a frequency of 77 Hz in
the linear regime.  Direct electric current (bias) was applied
simultaneously with AC excitation through the same current leads (see
insert to fig. 1).  Although we have studied, strictly speaking,
the differential resistance, for the sake of simplicity we will
refer to it below as resistance.   

\begin{figure}
\includegraphics{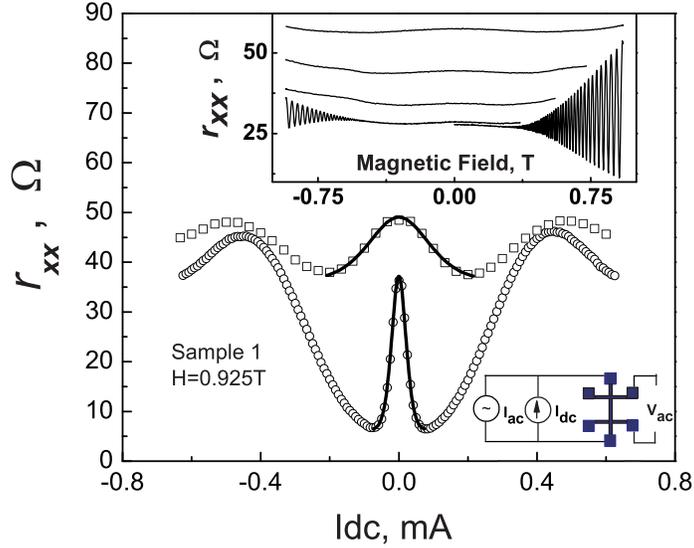}
\caption{\label{fig1}
Dependence of the differential resistance $r_{xx}$ on DC bias at H=0.925T.
Cirles correspond to T=4.3 K, squares correspond to T=19.8 K.  The solid
lines are  theoretical curves obtained from Eq. 3.  The fitting parameters
are $I_0$=0.055 (mA) and $\delta$=0.334 for T=4.3K and 
$I_0$=0.1802 (mA) and $\delta$=0.177 for T=19.8 K. The top inset shows
quantum oscillations of the longitudinal resistance at different
temperatures T=1.9 K (bottom curve, right), 4.2 K (bottom curve, left), 9.9, 19.8 and 35 K (remaining curves in ascending order).  The 
experimental set-up is shown at bottom right.  }
\end{figure}

Typical curves of the longitudinal resistance $r_{xx}$ as a
function of the DC bias are shown in Fig. 1 at two temperatures.  At
high DC bias the resistance exhibits maxima that satisfy the
condition $n\times\hbar \omega_c=2 R_c E_H$, corresponding to
"horizontal"  transitions between Landau levels\cite{yang,bykov1}.
Another striking feature is the sharp peak at zero DC bias which broadens
as the temperature is raised. This zero bias peak is the main topic of
our paper.  

\begin{figure}
\includegraphics{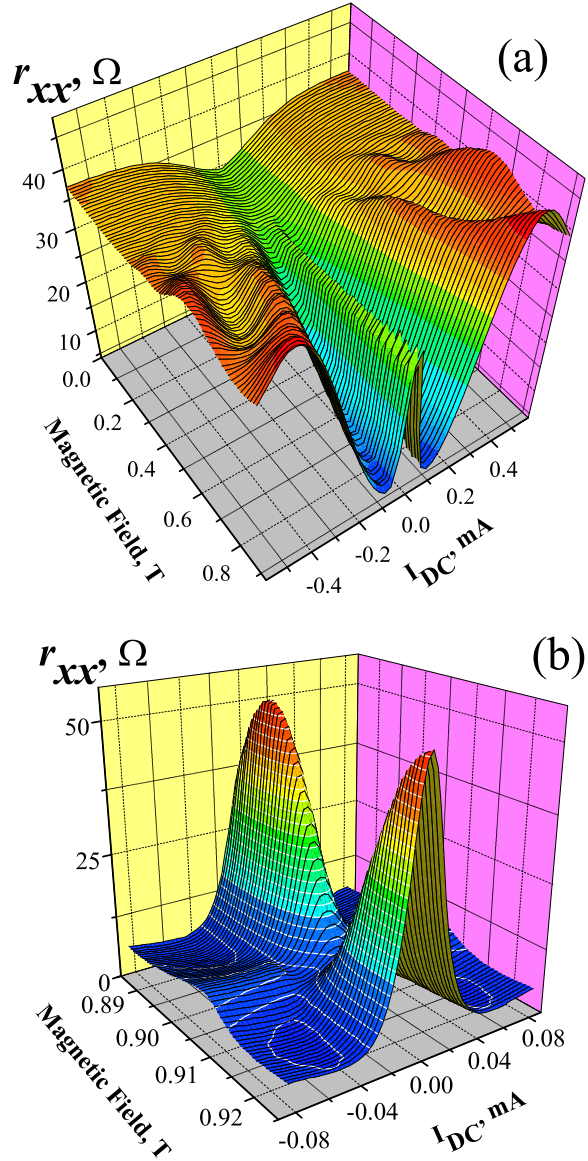}
\caption{\label{fig2}
(a) The differential resistance $r_{xx}$ as a function of magnetic field
and DC bias at temperature T=4.3 K.  (b) A similar plot at T=1.9 K
over a narrower range of the experimental parameters, yielding better
resolution. }
\end{figure}

The evolution of the magnetoresistance with DC bias and magnetic field is shown in Fig. 2.  The zero bias peak appears at relatively high
magnetic field $H \approx 0.2-0.3$ T (see fig.2a). At these fields the Landau level width $\hbar/\tau_q$ extracted from amplitude of Shubnikov de Haas (SdH) oscillations becomes to be comparable with $\hbar \omega_c $ and the SdH oscillations are visible at low temperatures (see curve at T=1.9K in the top insert to fig.1). The strength of the peak increases gradually with magnetic
field.  At the zero bias considerable SdH
oscillations are present in high magnetic field.  The magnitude of the SdH
oscillations at T=4.3 K is substantially smaller than the amplitude of the
zero bias peak.  The peak is still present at temperatures above T=30K
where no SdH oscillations are detected.  A better resolved snapshot of the
peak evolution is presented in Fig. 2b. The figure demonstrates the
effect  at low temperatures T=1.9 K, where the SdH oscillations are
well developed. 

The striking reduction of the resistance by several times is observed at
temperatures at which no SdH oscillations are present. This is quite
different from what one expects for electron heating by the electric
field.  As shown in the insert to Fig. 1, the resistance increases for
higher temperatures, in contrast with the observed decrease with applied
electric field.  It should also be noted that at low temperatures, the largest effect possible due to heating is to reduce the resistance from its value at a SdH maximum to the "average" baseline value (which is $\approx$26-28 ($\Omega$) in the insert to Fig. 1).  The observed reduction in Fig. 2b  is much greater than this indicating a new phenomenon associated with the application of an electric field\cite{resistance}.

From a theoretical perspective, nonlinear phenomena in high mobility 2D
electron systems can be conveniently separated into: (a) effects of
electric field on the electron distribution function \cite{dmitriev}, and 
(b) effects of electric field on the kinematics of electron scattering
\cite{durst,vavilov}.  It was recently realized that the the first of
these should provide the dominant contribution to the nonlinear response
in 2D electron systems.  Below we will compare our results with this
approach.\cite{dmitriev}

The theory considers 2D electrons in classically strong magnetic field at
finite electric field $E_{dc}$ and at relatively high temperature $T \gg
\hbar \omega_c$.  Due to conservation of total electron energy
$\epsilon+eE_{dc}x$ in the DC electric field $E_{dc}$, the spatial
electron diffusion translates into the diffusion of the electrons
in energy space.  The solution of the diffusion equation in
$\epsilon$-space yields nontrivial oscillations of the
nonequilibrium electron distribution function with period $\hbar
\omega_c$.  The amplitude of the oscillations is stabilized by inelastic
electron-electron scattering, which is found to be proportional to
$T^2$.  Relative to the Drude conductivity, $\sigma_D$, in zero magnetic
field, the theory predicts a longitudinal conductivity:

$$
\Delta
\sigma_{xx}/\sigma_D=2\delta^2[1-\frac{4Q_{dc}}{1+Q_{dc}}],  \eqno{(1)}
$$  
where $\delta=exp(-\pi/\omega_c \tau_q)$ is the Dingle factor, $\tau_q$ is
the quantum scattering time and the parameter $Q_{dc}$ is

$$
 Q_{dc}=\frac{2\tau_{in}}{\tau_{tr}}(\frac{eE_{dc}
v_F}{\omega_c})^2(\frac{\pi}{\hbar \omega_c})^2.    \eqno{(2)}
$$
Here $\tau_{in}$ is the inelastic relaxation time, $\tau_{tr}$ is the
transport scattering time and $v_F$ is the Fermi velocity.

In order to compare with experiment, the differential conductivity at
frequency $\omega$, $\sigma_{\omega}=dJ/dE=d(\sigma(E)E)/dE$, is obtained
using eq.(1), and the variation of the differential resistance is found
to be:
$$
 \Delta r_{xx}/R_0=2\delta^2[\frac{1-10Q_{dc}-3Q_{dc}^2}{(1+Q_{dc})^2}], 
\eqno{(3)}
$$ 
where $R_0$ is the resistance at zero magnetic field. 
In a classically strong magnetic field $\omega_c \tau_{tr} \gg 1$, the DC
electric field is almost perpendicular to the electric current $I_{dc}$:
$E_{dc}=\rho_{xy}I_{dc}/d$, where $d$ is the sample width.  Using Eq. 2,
we rewrite the parameter $Q_{dc}$ in the form $Q_{dc}=(I_{dc}/I_0)^2$,
where the scale $I_0$ is a fitting parameter.  In accordance with Eq. 3
the parameter $I_0$ is directly related to the width of the zero bias
peak and the peak  magnitude is proportional to $\delta^2$. 
Below we refer to the parameter $I_0$ as the linewidth. Examples of
theoretical fits to the data using Eq.3 are shown by the solid lines in
Fig. 1, using $\delta$ and $I_0$ as fitting parameters\cite{comparison}.

\begin{figure}
\includegraphics{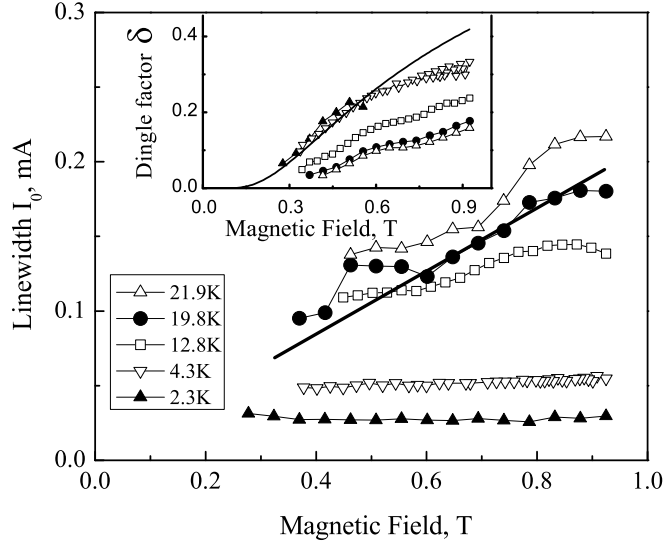}
\caption{\label{fig3}
Dependence of the width of the peak $I_0$ on  magnetic field 
at different temperatures, as labeled.  The solid line represents the
linear  dependence expected from the theory in the high temperature limit
(see Eq. 2).  The
inset shows the magnetic field dependence of the parameter $\delta$
obtained from the fit of the zero bias peak using Eq. 3.  The solid
line shows the theoretical dependence of the Dingle parameter
$\delta$ on magnetic field, corresponding to a quantum scattering time
$\tau_q$=1.5 (ps). }
\end{figure}

The dependence of the width of the peak ($I_0$) on magnetic field is
presented in Fig. 3 at different temperatures.  At high temperature
the peak width varies considerably with magnetic field. The
approximately linear increase of the scale $I_0$ with magnetic field 
agrees with the theory.  The deviations from the linear dependence
 are beyond the scope of the theory.  The
oscillations may be related to magneto-phonon resonances observed in
these systems at high temperature \cite{bykov2}.  At low temperature, a
regime that has not been considered by the theory, the width of the zero 
bias peak is found not to depend on magnetic field.  

For several different temperatures, the insert to Fig. 3 shows the
magnetic field dependence of the Dingle parameter, $\delta$, which is
obtained from comparison of the magnitude of the zero bias peak with the theory (see Eq. 3).
The parameter $\delta$ decreases with decreasing magnetic field, and disappears below
H=0.2T. Using theoretical expression for the Dingle parameter \cite{dmitriev} 
$\delta=exp(-\pi/\omega_c \tau_q)$, we have plotted the parameter $\delta$ vs magnetic field using the quantum 
scattering time $\tau_q$ as a fitting parameter. This is shown in the insert by the solid line.
The obtained quantum time $\tau_q=$1.5 (ps) is close to the quantum time extracted from
the usual analysis of SdH oscillations at different temperature
and/or magnetic field. A comparison between these two results is shown in the bottom insert to Fig. 4. 
Using this new method the time $\tau_q$ is found for temperatures up to 24K, where SdH
oscillations are not detectable and previous methods fail to work.
Thus the method extends considerably the temperature range, where
the  quantum scattering time can be studied.

\begin{figure}
\includegraphics{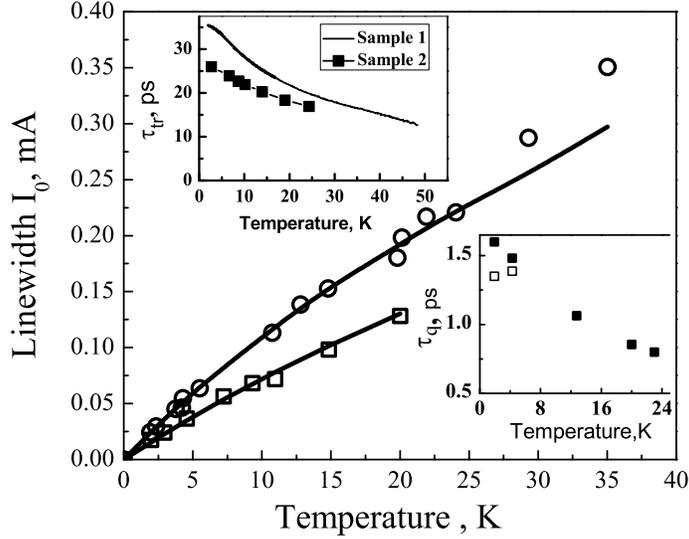}
\caption{\label{fig4}
Dependence of the width of the zero bias peak $I_0$ on temperature for
sample N1 (open cirles) and sample N2 (open squares) at H=0.925T.  The solid lines are
the theory using Eqs. 2 and 3.  The comparison gives an inelastic
scattering time $\tau_{in}=10/T^2$($12/T^2$)(ns) for sample N1(N2).    
The top inset shows the transport time $\tau_{tr}$ vs temperature at zero
magnetic field.  The dependence of the quantum scattering time $\tau_q$ on
temperature is shown in the bottom insert. Open squares correspond to
$\tau_q$ determined from the amplidute of the SdH oscillations using
Lifshits-Kosevich formulae\cite{ando}.  Filled squares are the $\tau_q$
determined by comparison of the amplidute of the zero bias peak with
Eq. 3. }
\end{figure}

The temperature dependence of the width of the peak is shown in Fig. 4 \cite{heating}. 
At low temperatures the width of the peak is found to be proportional to
the temperature $T$.  The linear temperature behavior of the $I_0$ indicates the
quadratic temperature dependence of the inelastic scattering time:
$\tau_{in}=\alpha/T^2$ (see eq.2), where $\alpha$ is a constant. This is in agreement 
with the theory\cite{dmitriev}. At higher temperature a noticeable sublinear deviation is observed. 
This can also be captured by the theory if the temperature variations of the transport
scattering time $\tau_{tr}$  is significant.  The temperature
dependence of $\tau_{tr}$ determined from the resistivity at zero
magnetic field is shown in the top insert to the figure. The solid lines in the 
main figure are theoretical curves plotted in accordance with  eq.2 and eq.3 in which 
the temperature variations of the $\tau_{tr}$ are taken into account    
and the constant $\alpha$ is the only fitting parameter.  For the inelastic time we have
found  $\tau_{in}^{(1)}=10 \times 10^{-9}/T^2$ (s) and $\tau_{in}^{(2)}=12 \times 10^{-9}/T^2$ (s) 
for sample 1 and 2 . The corresponding theoretical estimations
\cite{dmitriev} of the inelastic time give $\tau_{in,t}^{(1)}=4.8\times 10^{-9}/T^2$ (s) and
$\tau_{in,t}^{(2)}=3.2\times 10^{-9}/T^2$ (s). We consider this as satisfactory agreement,
in light of several approximations used in the theory.  The somewhat
larger values of the inelastic time $\tau_{in}$ obtained in the
experiment could also be due to additional electron screening
of 2D electrons by X-electrons in AlAs/GaAs type-II superlattices
\cite{fried}.  At high temperature $T>30$ K a considerable
super-linear temperature dependence of the linewidth is found (not shown
for sample N2).  This phenomenon has been left for a future study. 

In summary, a strong reduction of the longitudinal resitivity of 2D
electrons in classically strong magnetic fields is observed in response
to DC electric field.  We have found that the effect is not related to
Joule heating even at temperatures down to 2K, where strong quantum
oscillations (SdH) are present that are highly sensitive to the
temperature.  At low temperature (2-10K), the scale of the electric fields
at which the effect occurs is proportional to the temperature. Reasonable
agreement is established with recent theory \cite{dmitriev} that has
predicted significant and nontrivial variations of the electron
distribution function in response to a DC electric field. The
comparison with the theory allowed us to find inelastic electron-electron scattering time $\tau_{in}$ and the quantum scattering time $\tau_q$ of the 2D electrons at high temperatures where previous methods, based
on analysis of quantum oscillations, fail.

\begin{acknowledgments}

We thank prof. Myriam P. Sarachik for numerous discussions and technical
support.  S. V. thanks prof. Igor Aleiner for discussions.   This work was
supported by NSF : DMR 0349049 ; DOE-FG02-84-ER45153 and RFBR, project
No.04-02-16789 and 06-02-16869.

\end{acknowledgments}


\begin{references} 

\bibitem{zudov} M.A. Zudov, R. R. Du, J. A. Simmons, and J. L. Reno, Phys. Rev. B {\bf 64},201311(R) (2001); \bibitem{engel}P.D. Ye, L. W. Engel, D.C. Tsui, J. A. Simmons, J. R. Wendt, G. A. Vawter, and J. L. Reno, Appl. Phys.Lett {\bf 79},2193 (2001).
\bibitem{mani} R. G. Mani, V.Narayanamurti, K. von Klitzing, J. H. Smet, W. B. Jonson, and V. Umansky, Nature(London) {\bf 420}, 646 (2002)
\bibitem{zudov2} M.A. Zudov, R. R. Du, L. N. Pfeiffer, and  K. W. West, Phys. Rev. Lett {\bf 90} 046807 (2003).
\bibitem{dorozh1} S. I. Dorozhkin, JETP Lett. {\bf 77}, 577 (2003).
\bibitem{willett} R. L. Willett, L. N. Pfeiffer, and  K. W. West, Phys. Rev. Lett {\bf 93} 026804 (2004). 
\bibitem{andreev} A. V. Andreev, I. L. Aleiner, and A. J. Millis, Phys.Rev. Lett. {\bf 91}, 056803 (2003)
\bibitem{durst} A. C. Durst, S. Sachdev, N. Read, and S. M. Girvin, Phys. Rev. Lett. {\bf 91}, 086803 (2003)
\bibitem{anderson} P. W. Anderson and W. F. Brinkman, cond-mat/0302129
\bibitem{shi} J. Shi and X. C. Xie, Phys. Rev. Lett. {\bf 91}, 086801 (2003).
\bibitem{vavilov} M. G. Vavilov and I. L. Aleiner Phys. Rev. B {\bf 69}, 035303 (2004)
\bibitem{dmitriev} I. A. Dmitriev, M.G. Vavilov, I. L. Aleiner, A. D. Mirlin, and D. G. Polyakov, Phys. Rev. B {\bf 71}, 115316 (2005).
\bibitem{dorozh2}S. I. Dorozhkin, J. H. Smet, V. Umansky and K. von Klitzing Phys. Rev. B {\bf 71}, 201306(R) (2005).
\bibitem{yang} C. L.Yang, J. Zhang, and R. R. Du, J. A. Simmons and J. L. Reno, Phys. Rev. Lett. {\bf 89}, 076801 (2002)
\bibitem{bykov1} A. A. Bykov, Jing-qiao Zhang, Sergey Vitkalov, A. K. Kalagin, and A. K. Bakarov   Phys. Rev. B {\bf 72}, 245307 (2005)
\bibitem{resistance} The actual (not differential) resistance is 12.7 Ohm at DC bias $I_{dc}=$0.08 mA and H=0.92T in Fig.2b. This is considerably below the baseline value: average between resistances at maximum and nearest minimum ($\approx$26-28 $\Omega$). 
\bibitem{fried} K. J. Friedland, R. Hey, H. Kostial, R. Klann, and K. Ploog, Phys.Rev.Lett. {\bf 77}, 4616 (1996).
\bibitem{comparison} Although the theory  is developed in high temperature limit, we have used the formula (2) and (3) to find width of the peak and it's amplitude at all temperatures. Other fitting functions (procedures) do not change significantly the results presented in fig.3 and fig.4. 
\bibitem{ando} T. Ando, A. B. Fowler and F. Stern, Rev. Mod. Phys. {\bf 54}, 437 (1982). 
\bibitem{bykov2} A. A. Bykov, A. K. Kalagin, and A. K. Bakarov JETP Lett.{\bf 81}, 523 (2005). 
\bibitem{heating} Using energy relaxation rate $1/\tau_{\epsilon}$ measured in GaAs/AlGaAs (see "E. Chow {\it et al} Phys. Rev. Lett. {\bf 7}, 1143 1996) we have estimated the electron temperature at DC biases relevant to the zero bias peak. Negligibly small electron overheating by the DC biases is found:  $\Delta T \approx 0.1 K$ at T=4.2K and  $\Delta T \approx 0.01 K$ at T=10K

\end{references}

\end{document}